 \documentclass[prl,aps,twocolumn,showpacs]{revtex4}
\usepackage[dvips]{graphicx}
\usepackage{dcolumn}
\newcommand{\be}{\begin{equation}}
\newcommand{\ee}{\end{equation}}
\newcommand{\bea}{\begin{eqnarray}}
\newcommand{\eea}{\end{eqnarray}}
\newcommand{\nn}{ \nonumber}
\newcommand{\ds}{\displaystyle}
\begin{document}
\topmargin=-10mm

\title{Dynamical Kohn Anomaly in Surface Acoustic Wave 
Response in Quantum Hall Systems Near $ \nu = 1/2 $ } 

\author{Nataliya A. Zimbovskaya and Joseph L. Birman}
\vspace{2mm}

\affiliation{Department of Physics, The City College of CUNY, New York, NY, 10031, USA} 
 
\date{\today}

\begin{abstract} 
 The dynamical analog of the Kohn anomaly image of the Fermi Surface is demonstrated for the response functions  to the surface acoustic waves (SAW) in quantum Hall systems near $ \nu = 1/2. $ Kinks appear in the velocity shift $ \Delta s/s $ and attenuation coefficient $ \Gamma. $ The
effect is considerably enhanced under periodic modulation and should be observable in the geometry we propose, where the SAW wave vector is parallel to the wave vector of an external modulating potential.
              \end{abstract}
 \maketitle

In this paper we predict a new dynamical manifestation of the Kohn anomaly (KA) \cite{1} in the response of a two dimensional electron gas (2DEG), in the fractional quantum Hall effect (FQHE) regime to an interacting surface acoustic wave (SAW). 
This dynamical KA may be measurable from the SAW wave-vector dependence of the velocity shift and attenuation. The predicted kinks can be enhanced by an order of magnitude when external periodic modulation is applied, similar to the recent experiments \cite{2,3}. Our work is based on the HLR picture \cite{4}. So we assume that at $ \nu = 1/2 $ there exists a composite fermion-Fermi surface(CF-FS). For the unmodulated 2DEG the CF-FS is a circle \cite{5}.

 Surface acoustic wave propagation has been one of the most
fertile probes of the conducting properties of the 2DEG in high
magnetic fields, in particular, for the fractional quantum Hall effect (FQHE) in a 2DEG in the GaAl/AlGaAs heterostructures. 
Using the SAW mesurement, evidence was obtained for the existence of a CF-FS and the value of the Fermi momentum $ p_F $ was determined \cite{6}. The KA in the acoustic wave response is well known in conventional metals. It was noted by HLR (Ref. \cite{4}) that in principle a similar anomaly can also exhibit itself in the SAW response of the 2DEG at one-half filling. However, the authors concluded the effect would be very small. In this paper we show that in proper geometry and with modulation the effect can be sufficiently large to be measurable using the response of the system to the SAW at $ \hbar q \approx 2 p_F. $ And, if measured, this response would contain important information. It this paper we show that the image of the CF-FS appears in the structure of the SAW velocity shift $ \Delta s /s $ and the attenuation rate $ \Gamma, $ as functions of the SAW wave vector $ \bf q $ for large $ q \ (\hbar q \sim 2 p_F ) ,$  thus giving another means to determine the CF-FS.

The SAW propagating in a piezoelectric medium interacts with a 2DEG in a nearby semiconductor. In the usual geometry the SAW propagates along the [011] direction on the (100) surface of an Al$_x$Ga$_{1-x}$As crystal. Defining the $x$ axis as $ x \equiv [011] $ we have \cite{7}:

                 \begin{equation}
\frac{\Delta s}{s} - \frac{i \Gamma}{q} = \frac{\alpha^2}{2}
\varepsilon(q,\omega) = \frac{\alpha^2/2}{1 + i \sigma_{xx} (q,\omega)/\sigma_m}.
                 \end{equation}  
 Here $ \omega,{\bf q} $ are the SAW frequency and wave vector
$ ({\bf q} = (q,0,0); \  \omega = sq); $  the
dimensionless constant $ \alpha^2 $ describes the piezoelectric coupling, $ \sigma_{xx} (q,\omega ) $ is the component of the electron conductivity tensor, $ \ \sigma_m = s \varepsilon_b /2 \pi; \ \varepsilon_b $ is the background dielectric constant.
The dielectric function $ \varepsilon (q, \omega) $ is:
                 \begin{equation}
\varepsilon(q,\omega) = 1 - V(q) K_{00}(q,\omega ),
                 \end{equation}  
 where $ V(q) $ is the Fourier component of the Coulomb interaction: $ V(q) = 2\pi e^2/\varepsilon_b q; \ \varepsilon_b $ is the background dielectric constant; and $ K_{00} (q, \omega) $ is the CF density-density response function. Note that in Eq. (1) the dielectric function  $ \varepsilon(q,\omega) $ is in the numerator: this takes into account
that the piezoelectric field generated by the SAW is unscreened, while the 2DEG interacts with the screened electric field.

In the RPA the density-density response function is described by the following expression \cite{4}:
                 \bea &&
K_{00} (q, \omega) 
 \nn \\ \nn \\  &= &
 \frac{K_{00}^0 (q,\omega)}
{1 + V(q) K_{00}^0(q, \omega) -   (4 \pi \hbar /q)^2 K_{11}^0(q,\omega) K_{00}^0(q,\omega) }\,  . \nn \\ 
                 \eea  

Below we write the unperturbed response functions $
K_{11}^0(q,\omega) $ and  $ K_{00}^0(q,\omega) $ corresponding to the noninteracting CF system at $ \nu = 1/2, $ which is the Fermi sea with   Fermi  momentum $ p_F.$ The functions $ K_{00}^0 (q,\omega) $ and   $ K_{11}^0 (q,\omega) $  are:
                 \bea &&
K_{00}^0 (q,\omega) 
\nn \\  \nn \\ &=& 
\int \int \frac{f (E({\bf p} + \hbar{\bf q})) - f(E({\bf p}))}
{\hbar \omega - E({\bf p} + \hbar{\bf q}) + E({\bf p}) + i \eta}
\, \frac{d^2 p}{(2 \pi \hbar)^2} \, ,
         \eea
     \bea &&
K_{11}^0 (q,\omega)   \nn \\ \nn \\
 &=& - \frac{n}{m^*} +
 \int \int v_y^2 \frac{f (E({\bf p} + \hbar{\bf q})) - f(E({\bf p}))} {\hbar \omega - E({\bf p} + \hbar{\bf q}) + E({\bf p}) + i \eta} \, \frac{d^2 p}{(2 \pi \hbar)^2} \, .  \nn \\   
            \eea  
 Here $ f (E({\bf p})) $ is the Fermi distribution function for the energy $ E({\bf p}); \ v_y = p_y/m^*; \ m^* $ is the CF effective mass; $ \eta = \hbar/\tau; \ \tau $ is the relaxation time.

\section{Case A: Circular CF-FS}

After straightforward calculation we arrive at the result
$\ (\omega \ll q v_F; \ v_F = p_F/m^*; \  \eta \to 0): $
                 \bea
K_{00}^0 &=& N \left \{ 1 - 2 \delta \left[
\sqrt{\left (\frac{\hbar q}{2 p_F} -  \delta \right )^2 -1}
 \right. \right. 
 \nn \\ \nn \\ & +& \left. \left.
\sqrt{\left (\frac{\hbar q}{2 p_F} +  \delta \right )^2 -1}\; \right ]^{-1} \right \} .
                 \eea  
   where 
    \bea &&
 \sqrt{\left (\frac{\hbar q}{2 p_F} \pm  \delta \right )^2 -1}
\equiv
 \sqrt{\left| \bigg(\frac{\hbar q}{2 p_F} \pm \delta  \bigg)^2 -1\right |} 
 \nn \\ \nn \\ &\times& 
 \left \{ \theta
\bigg(\frac{\hbar q}{2 p_F} \pm \delta  -1 \bigg) + i \theta
 \bigg(\frac{\hbar q}{1 - 2 p_F} \mp \delta  \bigg) \right \} 
 \nn \eea
   and
     \[ 
 \theta (x) =
\left \{
\begin{array}{cc}
  0 \ \ \mbox{for} \  \ x \leq 0,  
 \nn\\ 
  1 \ \ \mbox{for} \  \ x > 0 . 
\end{array}
\right. 
 \]
    Here $ N = 2 \pi m^*/(2 \pi \hbar)^2 $ is the density of states at the CF-FS; $ \ \delta \equiv \omega/q v_F. $ Thus the unperturbed density-density response function exhibits an anomaly near $ \hbar q = 2 p_F $ both in  real and
imaginary parts. The main part of the response function 
$ K_{11}^0 (q,\omega) $ at $ \omega \ll q v_F \ (\delta \ll 1) $ equals 
                 \bea 
K_{11}^0  &=& -\frac{q^2}{24 \pi m^*} - \frac{2}{3} \frac{n}{m^*} \frac{k_F}{q}  \left \{ 
 \sqrt{\left[ \bigg(\frac{\hbar q}{2 p_F} -\delta  \bigg)^2 -1\right ]^3} \right.
   \nn \\ \nn \\ & &-   \left.
\sqrt{\left[ \bigg(\frac{\hbar q}{2 p_F} + \delta  \bigg)^2 -1\right ]^3} \,\right \}.
                 \eea  
   For small $ q \ (\hbar q \ll p_F ), $ Im$K_{11}^0 $ is large relative to the real part and gives the linear dependence of the conductivity $ \sigma_{xx} $ upon $ q. $ However, for $ \hbar q \sim p_F $ and $ \sigma \ll 1, $ Im$K_{11}^0 $ is small compared to Re$ K_{11}^0. $ The ratio of these magnitudes is of the order of $ \delta. $   

Substituting the expressions (6) and (7) in Eq. (3) and using Eq. (1) we obtain the following result for the $ \sigma_{xx} $ component of the electron conductivity tensor:
                 \bea &&
\sigma_{xx} 
\nn \\ \nn \\ &=& \!\!\! - i \sigma_0(q) \nn \\ \nn \\ \! &\times& \!
\left [ 1 - \frac{3}{2} \frac{\delta}{\ds
\sqrt{\bigg (\frac{\hbar q}{2 p_F} -  \delta \bigg )^2\! -1} +
\sqrt{\bigg (\frac{\hbar q}{2 p_F} +  \delta \bigg )^2 \!-1} 
-  \frac{\delta}{2}} \right ]
 \nn \\ \nn \\ 
  & + & \!\frac{3}{4} \delta \sigma_0 (q) 
   \nn \\ \nn \\ \!\!\! &\times& \!
\left [ 1 +\frac{1}{2} \frac{\delta}{\ds
\sqrt{\bigg (\frac{\hbar q}{2 p_F} -  \delta \bigg )^2\! -1} +
\sqrt{\bigg (\frac{\hbar q}{2 p_F} +  \delta \bigg )^2 \!-1} 
-  \frac{\delta}{2}} \right ] , \nn \\
                 \eea 
where $ {\sigma_0(q) = \frac{3}{4} (\omega e^2/q^2) N.} $ 

In the vicinity of $ \hbar q = 2 p_F $ the denominators are of the order of $ \delta, $ therefore the last two terms are smaller than the first two terms. Neglecting them, we obtain for the real part  $ (\sigma') $ of  $ \sigma_{xx}\ $
                 \begin{equation}
\sigma' = \sigma_0(q) F(q).
                 \end{equation}
The function $ F(q) $ increases when the wave vector approaches   $ 2 p_F/\hbar $ (but $ 1 - \hbar q/2 p_F \gg \delta) $ as   
                 \begin{equation}
  F(q) = \frac{3}{8 \sqrt2}\, \frac{\delta}{\sqrt{1 - \hbar q/2 p_F}}
     \ee
  and  has its maximum value ${ F_{\max} \ \big (F_{\max} =  \frac{3}{16} \sqrt \delta \big) } $ when $ \hbar q = 2 p_F(1 - \delta).$ Further increase of the wave vector leads to  decrease of the function  $ F(q) $ which goes to zero when $ \hbar q > 2 p_F (1 + \delta). $

The derivative of the function $ F(q) $ tends to infinity
when $ \hbar q $ approaches the points $ 2p_F(1 - \delta) $ and 
$ 2 p_F(1 + \delta) $ from the left side. These peculiarities in  $ F'(q) $ correspond to kinks in the dependence of  the real part of the conductivity upon $ q $ near the ''KA point'' $ (\hbar q = 2 p_F). $ Due to the low density of electrons in the heterostructures under consideration $ (n \sim 10^{10} - 10^{11} $ cm$^2) $ the CF  Fermi velocity is comparatively small $(v_F \sim 10^6 $ cm/sec). So the parameter  $\delta $ cannot be  neglected $ (\delta \sim 10^{-1}; \ s \approx 3 \cdot 10^5 $ cm/sec \cite{5}) and the variations of $ \sigma' $ near $ \hbar q = 2 p_F $ can be important. The imaginary part $ \sigma''$  of $\sigma_{xx} $ is not very sensitive to the variation in $ q $ in the vicinity of the point $ q = 2 p_F/\hbar, so $
                 \begin{equation}
\sigma'' \approx - \sigma_0 (q).
                 \end{equation}
  However, the derivative of this function diverges when $ \hbar q $ approaches the points $ 2 p_F(1 \mp \delta) $
from  the right side. This corresponds to  the results obtained in long ago \cite{9} for a metal with a nearly cylindrical Fermi surface. The real and imaginary part of the dielectric function are
                 \be 
\frac{\Delta s}{s}   =  \frac{\alpha^2}{2}
\mbox{Re} \varepsilon (q, \omega = s q) 
  =  \frac{\alpha^2}{2}
 \frac{1}{1 + \overline\sigma^2}
 \frac{1}{\displaystyle{1 +  \frac{\overline \sigma^2 }{1 + \overline\sigma^2} F^2 (q)}} , \ee
                \be
\Gamma = - q \frac{\alpha^2}{2}
\mbox{Im} \varepsilon (q, \omega = s q) = 
q \frac{\alpha^2}{2}  \frac{\overline \sigma}{1 + \overline\sigma^2} \frac{F(q)}{\displaystyle{1 +  \frac{\overline \sigma^2 }{1 + \overline\sigma^2} F^2 (q)}} ,
                 \ee
 where $\overline \sigma = \sigma_0(q) /\sigma_m. $
These formulas predict a minimum in the velocity shift and a maximum in the attenuation at  $ \hbar q = 2 p_F (1 - \delta) $ \cite{10}. Thus when the CF-FS is a circle the KA appears as non-monotonic behavior of the SAW velocity shift and attenuation dependencies of $ q $ near $ q = 2 p_F/\hbar. $
Equations (12) and (13) are our new results and are illustrated
on Fig. 1. Note that the location of the anomalies are shifted from $ 2p_F$ \cite{9}.

\begin{figure}[t] 
\begin{center}
\includegraphics[width=5.0cm,height=4cm]{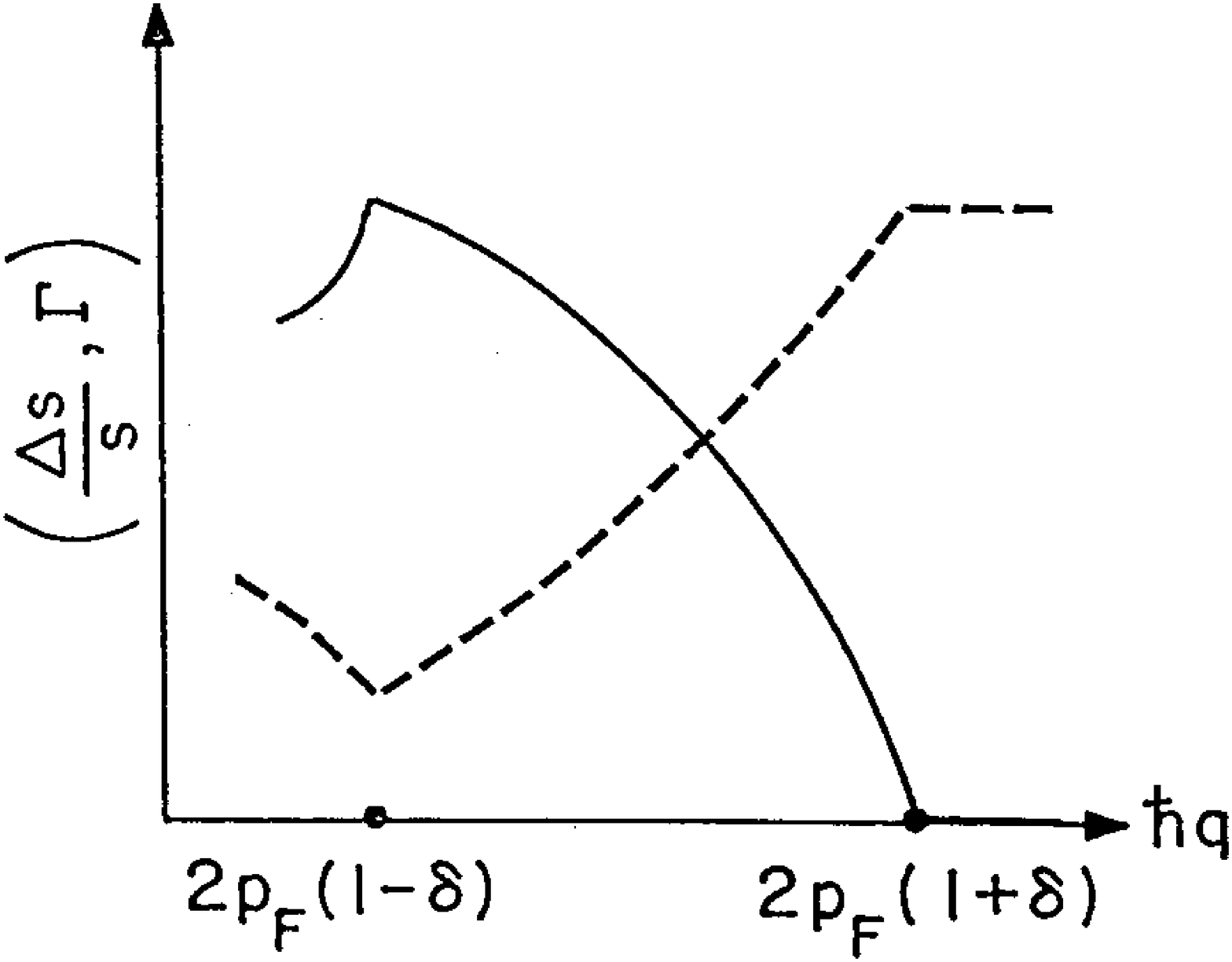}
\caption{The dependence of the SAW velocity shift $\Delta s/s $ (dashed curve) and attenuation coefficient $ \Gamma $ (solid curve) upon $ q $ near the KA point for case A, circular CF-FS. See text.
}   
\label{rateI}
\end{center}
\end{figure}

The ratio $ \sigma_0(q)/\sigma_m $ for GaAs at $ q \approx 2 p_F/\hbar $ and $ n \sim 10^{10}$ cm$^{-2} $ is of the order of unity and the piezoelectric constant $ \alpha^2 $ is of the order of $ 10^{-3} $ \cite{8,11}. So the magnitude of the minimum of the SAW velocity shift  can be of the order of $ 10^{-4} - 10^{-5}. $ According to HLR \cite{4}, a more systematic calculation of the response functions at large $ q / (q \sim 2 p_F /\hbar ) $ including non-RPA contributions gives a significant suppression of the KA due to strong interaction between electrons in the 2D system. Estimations by HLR (Ref. \cite{4}) gave an estimate of $ 0.1 $ reduction. Taking this estimate into account, we conclude that the dynamical KA in  the SAW response of the 2DEG with the undeformed (circular) CF-FS is probably not large enough for observation in experiments.

\section{case B: Deformed CF-FS}

\begin{figure}[t] 
\begin{center}
\includegraphics[width=9.3cm,height=7cm]{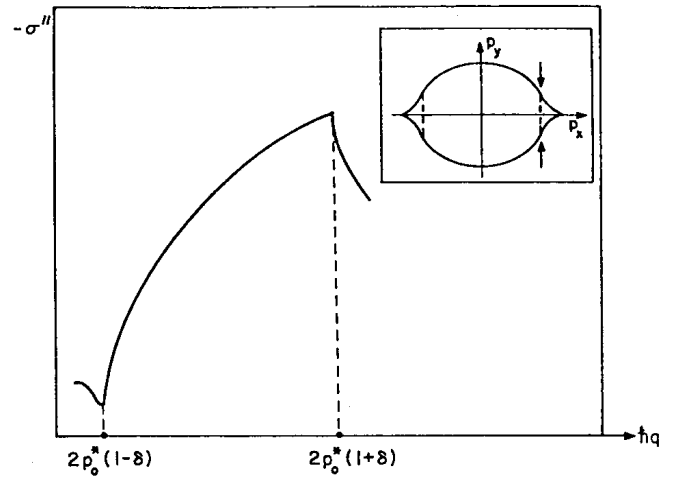}
\caption{The dependence of $ \sigma'' $ on $ q $ in the vicinity of the KA points for case B, Deformed CF-FS. A strongly flattened CF-FS is illustrated $ (\gamma \gg 1). $ See text. The inset shows schematically the shape of the CF-FS deformed due to an external periodic modulation in the ''x'' direction in the nearly free electron approximation (solid line) and CF-FS corresponding to Eq. (16) (dashed
line). The locations of regions of particularly high density of states are noted by two arrows (symmetrical points exist by reflection).
}   
\label{rateI}
\end{center}
\end{figure}

We predict a much stronger KA in the SAW
spectrum when the CF-FS is deformed due to an external effect.
For instance, the CF-FS can be flattened as a result of the
application of a periodic modulation potential, as in the experiments \cite{2}. The modulation will influence the CF system in two ways: through the direct effect of the modulating potential and through the effect of the inhomogeneous correction to the effective magnetic field $ \Delta B (r) $ proportional to the density modulation $ \Delta n (r). $ The direct effect of the modulation is that the modulating electric field deforms the CF-FS; this is analogous to the crystalline field in metals. As a result, in the presence of the grating modulation the CF-FS circle is distorted and can be "flattened"in the vicinities of some special points (see Fig. 2). It is known that for large $ q, $ resonance mechanisms of the absorption of the energy of SAW by the CFs, independent of quasiparticle scattering, are the most important. In this case the main contribution to the response functions originates from small effective parts of the CF-FS, where CFs which are strongly interacting with the SAW are concentrated. When the locally "flat" regions of the CF-FS are its "effective parts', they play a disproportionately important role in determining the SAW response due to the unusually large density of quasiparticle states there. We showed in our  work \cite{12} that this deformation of the CF-FS resulting from the modulating potential could be at the origin of the anomalies in magnetotransport observed in Ref. \cite{2}. Here we show that the distortion of the CF-FS can essentially strengthen the KA and make it large enough for observation in experiments.

Assume that the modulation potential wave vector $ \bf g $ is parallel to the SAW wave vector $ \bf q. $ This parallel geometry is the most favorable for our new experiment under deformation because, as in a conventional metal, it provides the  "flattened", segments of the CF-FS which are its effective parts for the absorption of phonons with large $ q $ of the order of $ 2 p_F/\hbar. $ \cite{13}. To analyze the Kohn effect under these conditions we parametrize the energy-momentum relation for CF in the modulated system  as in our previous work \cite{12}:
                 \begin{equation}
E({\bf p}) = \frac{p_x^2}{2 m_1} + \frac{p_0^2}{2 m_2}
\left |\frac{p_y}{p_0} \right|^\gamma
                 \end{equation}  
  where $ p_0 $ is a constant with the dimension of momentum, the $ m_i $ are effective masses and $ \gamma $ is a dimensionless parameter which determines the shape of the CF-FS $ (\gamma > 1). $  When $ \gamma > 2 $ the CF-FS is flattened near the vertices $ (\pm p_0,0). $ The CF-FS will be the flatter
at $(\pm p_0, 0) $ the larger is the parameter $\gamma  $.
To calculate the response functions $ K_{00}^0 $ and  $ K_{11}^0 $ we introduce
                 \begin{equation}
p_x =  p_0 \sqrt{m_1/m_2}\, \cos t, \qquad 
p_y = \pm p_0 |\sin t|^{2/\gamma} 
                 \end{equation}  
  where $ 0 \le t \le 2\pi $ and the $ +$ and $ - $ signs are chosen corresponding to normal domains of positive and negative values of the sine. To proceed we convert the integral in (6) to an integral over the CF-FS and over energy.

The expression (4) becomes:
                 \begin{equation}
K_{00}^0 (q,\omega) = \Phi_+(q,\omega) - \Phi_-(q,\omega),
                 \end{equation}  
  where
                 \bea
\Phi_\pm (q, \omega) &=& \frac{8 b_\pm}{\gamma} 
\frac{m_1 m_2}{\hbar q p_0} \frac{1}{(2 \pi \hbar)^2}
   \nn \\ \nn \\ & \times&
\int d E f (E) \int_0^{\pi/2}
\frac{\sin t^{2/\gamma - 1} d t}{(b_\pm)^2 - \cos^2 t}\, ;
                 \eea 
$ \displaystyle{b_\pm = \hbar q/2 p_0^* \mp \delta + i/q l; \ p_0^* = \sqrt{m_1/m_2}\,p_0 } $ is the maximum value of the component of the CF momentum in the direction of the SAW
propagation; $ \delta = \omega/q v_0; \ v_0 = p_0\big /\sqrt{m_1 m_2}; \ l = v_0 \tau. $

The inner integrals in Eq. (17) can be transformed to:
   \[\ 
\frac{1}{2 b_\pm^2} \int_0^\infty z^{1/\gamma - 1} 
(1 + z)^{1/\gamma - 1/2} \left( z + \frac{b_\pm^2 - 1}{b_\pm^2}
\right)^{-1} dz 
    \]
            \be       
= \frac{1}{2 b_\pm^2} B \left (\frac{1}{\gamma}; \frac{3}{2} -
\frac{2}{\gamma} \right ) {_2F_1} \left ( 1; \frac{1}{\gamma};
\frac{3}{2} - \frac{1}{\gamma}; 1 - \frac{b_\pm^2 - 1}{b_\pm^2}
\right). 
                 \ee 
Here B(x,y) is the beta function and $ _2 F_1 (\alpha; \beta; \rho; x)$ is the hypergeometric function. Using the asymptotic expression for the hypergeometric function in the limit $ |b_\pm^2 - 1| \ll 1, $ we obtain the following
expression for the response function $  K_{00}^0 (q,\omega) $
near $ q = 2 p_0^* (\delta \ll 1; \ \eta \to 0): $
            \bea 
K_{00}^0 (q,\omega) &=& N \left \{1 + \left(\frac{p_0^*}{\hbar q} \right)^{2/\gamma} 
\left [\left(\frac{\hbar q}{2 p_0^*} - \delta \right )^2  - 1 \right
]^{1/\gamma}  \right.
              \nn \\ \nn \\ &- &
\left.  \left(\frac{p_0^*}{\hbar q} \right)^{2/\gamma} 
\left [\left(\frac{\hbar q}{2 p_0^*} + \delta \right )^2  - 1 \right ]^{1/\gamma} \right \} 
  \nn \\ \nn \\ &
\equiv& N \{1 + Y_\gamma (q,\delta) \} ,
                 \eea 
where the density of states $ N $ equals $ \displaystyle{
2\pi \sqrt{m_1 m_2} 2^{2/\gamma}/\gamma \sin (\pi/\gamma) (2
\pi \hbar)^2} . $ 

Similarly  we can obtain the asymptotic expression for
the response function $ K_{11}^0(q, \omega) $ near the point of the anomaly. The expression for the $ \sigma_{xx} $ component of the electron conductivity tensor for the  flattened CF-FS is
                 \begin{equation}
\sigma_{xx} = - i \sigma_0 (q) \left \{ 1 -\frac{Y_\gamma(q, \delta)}{1 + a + Y_\gamma(q, \delta)} \right \} ,
                 \end{equation}  
 where $ \displaystyle{\sigma_0(q) = (\omega e^2/q^2) N/(1+a)};\
$ the dimensionless constant ''a''  equals $ 4^{1/\gamma-1}
\gamma/(2\gamma - 1), $ and ''a'' carries the information about the ''flattening'' of the initially circular CF-FS.

Equation (21) is our penultimate result and from it we obtain
important features of the KA for the deformed CF-FS.
When $ \gamma = 2 $ and $ m_1 = m_2 = m^* $ (undeformed CF-FS)  this formula will coincide with Eq. (10). 
Near $ \hbar q = 2 p_0^* \  [p_0^* $ is defined below Eq.(17)], 
for strong flattening of the CF-FS $ (\gamma \gg 1) $ the real part $ \sigma' $ of the conductivity $ \sigma_{xx} $ is very small compared to $ \sigma_0(q) $ and can be neglected.
Its imaginary part $ (\sigma'') $   near but not too close to 
$ \hbar q = 2 p_0^* \  (\hbar q < 2p_0^*) $ equals $ -\sigma_0(q) $ as in the case when the CF-FS in undeformed.
 
The imaginary part of the conductivity at $ \hbar q \approx 2 p_0^*(1 - \delta) $ is now
                 \begin{equation}
\sigma'' = - \sigma_0 (q) \left \{ 1 - 
\delta^{1/\gamma} \, \frac{\cos(\pi/\gamma)}{1 + a + \delta^{1/\gamma} \cos(\pi/\gamma)} \right \}.
                 \end{equation}  
 Its magnitude decreases significantly in the vicinity of
the KA point when CF--FS flattening is strong $ (\gamma \gg 1). $ When  $ \hbar q $ tends to
$ 2p_0^*(1 + \delta)\ [\hbar q < 2 p_0^* (1 + \delta)] $ the quantity $ \sigma''$ increases sharply, and
                 \begin{equation}
\sigma'' \approx - \sigma_0(q) \left \{1 + 
\delta^{1/\gamma} \left [1 + a -  \delta^{1/\gamma}
\right ]^{-1} \right \}.
               \end{equation}  
 When $ \gamma \gg 1 $ the parameter $ \delta^{1/\gamma} $ is of the order of unity and $ \sigma' \sim - \sigma_0 (q)(1 + a)/a. $ Then for strong flattening of the CF-FS the factor $ (1 + a)/a $ varies from 1 to 10.

The wave-vector dependence of the function $ \sigma''(q) $ near the point  $ \hbar q = 2 p_0^* $ is shown in Fig. 2. We observe kinks at points $ \hbar q = 2 p_0^*(1 \mp \delta) $ corresponding to discontinuities in the derivative of the imaginary part of the conductivity. Thus, flattening  the CF-FS  will critically  change the dependence of $ \sigma $ on
$ q $ in the vicinity of the point $ \hbar q = 2 p_0^*. $ In this case the most important contribution to the  Re$ \varepsilon (q, \omega = q s) $ is  
                 \begin{equation}
\mbox{Re} \varepsilon (q, \omega = q s) = 
\frac{1}{1 + |\sigma''/\sigma_m|} \,       
           \end{equation}  
 and, correspondingly
                 \begin{equation}
\frac{\Delta s}{s} = \frac{\alpha^2}{2} \,
\frac{1}{1 + |\sigma''/\sigma_m|} \,
.              \end{equation}  

For $ \gamma \gg 1, \ \Delta s/s $ will decrease significantly in the region $ 2 p_0^*(1 - \delta ) < \hbar q < 2 p_0^*(1 + \delta ).  $ The amount of decrease in Re$ \varepsilon(q,\omega) $ may be of the order of 10. Thus the anomaly produces a significant supression of the piezoelectric coupling (i.e. enhanced screening of the electric field) by the electrons at $ 2 p_0^*(1 - \delta ) < \hbar q < 2 p_0^*(1 + \delta ). $
Further increase of $ q $ leads to a sharp increase in $ \Delta  s/s , $ which corresponds to the weakening of the electron systems coupling to the SAW. This follows because the electrons can not absorb phonons with wave vectors more than $ 2p_F(1 + \delta)/\hbar. $ The  magnitude of this dip in the SAW
velocity shift for a strongly flattened CF-FS can be of  the order $ 10^{-3}. $ Even estimating as did HLR (Ref. \cite{4}) the reduction of the magnitude of the effect due to the contribution from non-RPA terms, we can conclude (using the HLR estimate \cite{4} of 0.1 reduction) that the predicted effect
should be observable by presently available experimental methods in modulated 2DEG systems.

In summary, we showed that  dynamical KA can be observed in the SAW velocity shift and attenuation in modulated 2DEG near $ \nu = 1/2. $ The distortion of the CF-FS by the external density modulation produces an order of magnitude enhancement and can make the anomaly available for observation in spite of the suppression due to the electron-electron interaction in 2D systems. The predicted shifts in location of the kinks from $ 2 p_F $ is also new. We propose an SAW experiment in a modulated 2DEG to observe this dynamical KA. We expect it to be observable in the $GaAl/Al_x Ga_{1-x} As $ heterostructures with the density of electrons $ n \sim 10^{10}$ cm$^{-2} $ at the SAW frequency $ \omega \sim 2 \pi (1-10) $GHZ in the geometry when the SAW wave vector is directed along the wave vector of modulating potential.

\vspace{2mm}

{\it  Acknowledgments:}
We thank Dr. R.L. Willett and Dr. S.H. Simon for discussions and
Dr. G.M. Zimbovsky for help with the manuscript. 
This work was supported in part by a grant from the
National Research Council COBASE Program.


\begin{references}
\bibitem{10} A slight modification of this statement is needed. The actual location of the minimum (maximum) in $ \Delta s/s \; (\Gamma) $ is slightly closer to $ \hbar q = 2 p_F. $ This point can be found by taking the $ q$-dependence of $ \sigma'' $ into account more carefully, near $ \hbar q = 2 p_F. $
\end{references}
\end{document}